\documentclass[prb, aps, twocolumn, floatfix, 10pt]{revtex4-1}

\usepackage{graphicx}
\usepackage{epstopdf}

\begin{document}

\title{Nonlinear microwave response of aluminum weak-link Josephson oscillators}
\author{E. M. Levenson-Falk}
\author{R. Vijay}
\author{I. Siddiqi}
\email[Corresponding author: ]{irfan@berkeley.edu}
\affiliation{Quantum Nanoelectronics Laboratory, Department of Physics, University of California, Berkeley CA 94720}

\date{\today}

\begin{abstract}

We present the driven response at T=30mK of 6 GHz superconducting resonators constructed from capacitively-shunted three dimensional (3D) aluminum nanobridge superconducting quantum interference devices (nanoSQUIDs). We observe flux modulation of the resonant frequency in quantitative agreement with numerical calculation and characteristic of near-ideal short weak link junctions. Under strong microwave excitation, we observe stable bifurcation in devices with coupled quality factor ($Q$) ranging from $\sim 30-3500$. Near this bias point, parametric amplification with $>$ 20dB gain, 40 MHz bandwidth, and near quantum-limited noise performance is observed. Our results indicate that 3D nanobridge junctions are attractive circuit elements to realize quantum bits.

\end{abstract}


\maketitle

A Josephson junction\textemdash two superconducting electrodes separated by a barrier\textemdash operated below its critical current $I_{0}$ and transition temperature $T_{C}$ behaves as a nonlinear inductor with vanishing internal dissipation. Junctions can thus be used to construct low-loss anharmonic oscillators--key components of many quantum bits (qubits) and sensitive amplifiers.  In particular, junctions based on nanoscale metallic weak links rather than tunnel barriers can avoid the 1/f noise and dissipation attributed to amorphous insulating layers\cite{Martinis}, potentially reducing qubit decoherence.  Furthermore, such narrow constrictions are, by geometry, well-suited for efficient magnetic coupling to nanoscale spin ensembles such as magnetic molecules \cite{Wernsdorfer_review, CSIRO, Hilgenkamp} and doped semiconductors \cite{Awschalom PNAS review}, and are thus attractive for realizing hybrid quantum information processing systems.  The electrical properties of a weak-link junction depend strongly on the dimensions of the constriction and adjacent structures\cite{Likharev}.  In a three-dimensional topology with thick banks connected by a thin, narrow bridge, the current-phase relation (CPR) can approach the theoretical limit of a short wire with fixed phase reservoirs at the ends, with nonlinear transport properties similar to those of a tunnel junction \cite{vijay_wire, vijay_apl}. 

In this letter, we report the RF driven response of anharmonic resonators based on 3D aluminum nanobridge Josephson junctions at $T \ll T_C$, where quasiparticle loss is minimized and high internal quality factors ($Q_{\mathrm{int}}$) can be achieved.  We investigate circuits designed with external $Q$ ranging from 30-3500. All devices exhibit strong nonlinearity, evidenced by stable bifurcation and high-gain, near-quantum-noise-limited parametric amplification. Our results are quantitatively consistent with the CPR of a short 3D weak link junction, which has been numerically computed \cite{vijay_wire} and experimentally verified \cite{vijay_apl}. These devices are thus a promising circuit element for realizing high-performance Josephson qubits and amplifiers.

\begin{figure}
\includegraphics{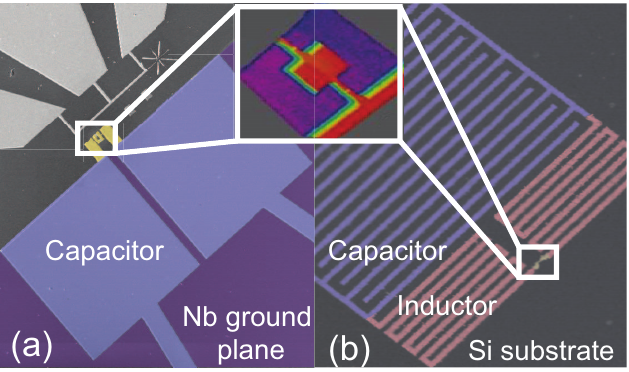}
\caption{\footnotesize\label{fig1} (color online) (a) Falsecolor SEM image of a typical low-$Q$ resonator.  The pads over the Nb ground plane form two parallel-plate capacitors in series (total C = 7pF);  these shunt a nanoSQUID which provides the inductance (total L = 80pH). An AFM image of a sample nanoSQUID is inset.  (b) Falsecolor SEM image of a typical high-$Q$ resonator.  A meander inductor is interrupted by a nanoSQUID, giving total L = 1.25nH; this is shunted by an 0.53pF finger capacitor.}
\end{figure}  

We study two resonator geometries, both employing a superconducting quantum interference device (SQUID) consisting of two 100 nm long, 30 nm wide aluminum nanobridges on a loop. The SQUID behaves effectively as a single junction with a magnetic-flux-tunable Josephson inductance. This allows us to vary the resonant frequency ($f_{\mathrm{res}}$) in the 4-8 GHz band, an important functionality for practical circuits, and to correlate the observed RF response with the expected CPR, complementing recent studies of single-junction weak link resonators\cite{Bezryadin_wire}. Our first device is designed for low $Q$ and consists of a nanoSQUID shunted by two parallel-plate capacitors (Fig. 1a) with a SiN dielectric. The resonator is directly coupled to a  $50\Omega$ transmission line via a $180^o$ hybrid to obtain $Q=30$. Our second device uses interdigitated finger capacitors to avoid the loss associated with the SiN layer and thus achieve high $Q_{\mathrm{int}}$. However, this structure's capacitance is too small to achieve the same $f_{\mathrm{res}}$; we thus place a meander inductor in series with the nanoSQUID (Fig. 1b), boosting the inductance at the expense of diluting the nonlinearity of the Josephson element. Coupling capacitors isolate the resonator from the $50\Omega$ transmission line, providing a total $Q = 3500$. Devices were patterned on bilayer resist by electron-beam lithography. Double-angle evaporation with a standard liftoff process was used to deposit the bridge and contacts (8nm and 90nm thick, respectively).  Measurements were taken in a dilution refrigerator at $T \approx $ 30mK $<< T_c$.  Cryogenic filters, isolators, and low-noise amplifiers were used to perform microwave reflectometry in the 4-8 GHz band as a function of drive frequency, drive power, and magnetic flux.

All devices exhibit a typical, linear resonance at low input power ($\approx -105$ dBm) and zero net flux.  By fitting the resonance curves for the high-$Q$ device, we extract $Q_{\mathrm{int}}$ as a function of power.  At a resonator photon occupancy of $\overline{n} \approx 1$, $Q_{\mathrm{int}} \approx 50000$.  We observe that $Q_{\mathrm{int}}$ rises with increased drive power, consistent with loss due to a bath of two-level systems (TLS), perhaps on top of the metal films or at metal-insulator interfaces\cite{Martinis_TLS, Clarke}.  Identical resonators without nanoSQUIDs show similar $Q_{\mathrm{int}}$, including the power dependence. We do not observe performance degradation due to the weak-link even as the drive power is increased, which might arise if phase slippage or other loss mechanisms were present.   

\begin{figure}
\includegraphics{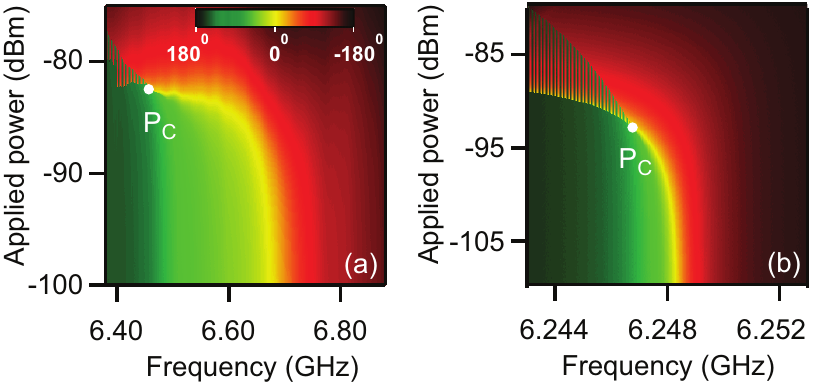} 
\caption{\footnotesize\label{fig2} (color online) Power dependence of low-$Q$ (a) and high-$Q$ (b) resonators.  At low power, a linear resonance is observed.  At higher powers, $f_{\mathrm{res}}$ decreases and the resonance becomes nonlinear.  Above a critical power $P_C$, the resonator bifurcates into two stable states.  In this regime the response is hysteretic in drive power; traces sweeping power up and down are interlaced in the plots, leading to striping in the hysteretic regime.}
\end{figure}

At microwave drive powers approaching the critical power $P_{C}$, we access the nonlinear regime (Fig. 2).  We plot the phase of the reflected signal, ranging from $+180^o$ below resonance to $-180^o$ above resonance, with $f_{\mathrm{res}}$ at $0^o$ (indicated in yellow).  The data show a linear resonance at low power.  As drive power increases, the resonance becomes nonlinear\textemdash the phase shift becomes sharper and asymmetric\textemdash and $f_{\mathrm{res}}$ decreases.  At powers of $P_{C}\approx -90$ dBm and $P_{C}\approx -85$ dBm for the high-$Q$ and low-$Q$ samples, respectively, the resonance bifurcates into two stable states with different oscillation amplitude.  This is characteristic of an anharmonic oscillator with a softening potential \cite{landau-lifshitz,Vijay_thesis}. After bifurcation, the resonator switches hysteretically between the two states depending on the direction of power sweep, as shown by the vertically interlaced traces in Fig. 2.  Using the Duffing model and the frequency of the critical point, we can extract the $Q$ of the resonator, giving $Q = 29$ and $Q = 3400$ for the low-$Q$ and high-$Q$ devices, respectively, in excellent agreement with our measurements of the linear resonance linewidth. The bifurcation regime is stably accessible even in the low-$Q$ device, thus indicating strong nonlinearity in the nanobridge junctions.

\begin{figure}
\includegraphics{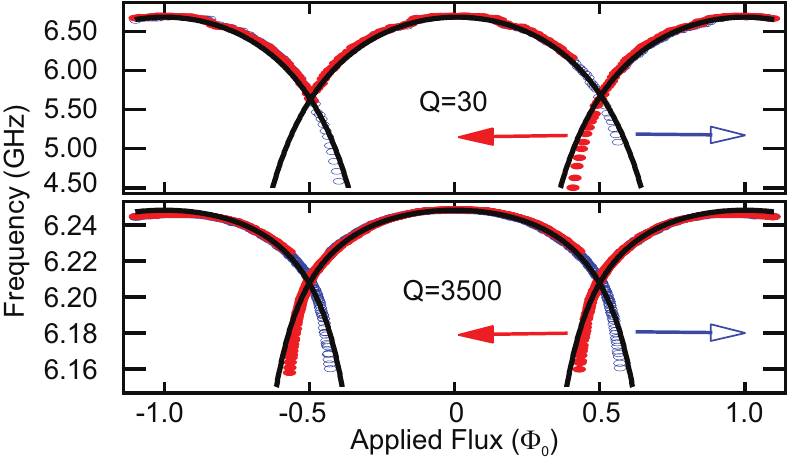} 
\caption{\footnotesize\label{fig3} (color online) Resonant frequency ($f_{\mathrm{res}}$) as a function of applied magnetic flux the nanoSQUID for the low-$Q$ (a) and high-$Q$ (b) resonators, showing $\Phi_0$-periodic modulation.  Data taken with flux swept negative to positive is shown as unfilled blue circles; positive to negative, filled red circles.  Hysteresis is observed past half-integer flux quanta.  Theoretical modulation curves using a numerically calculated CPR are plotted as solid black lines.}
\end{figure}

We can tune $f_{\mathrm{res}}$ by applying a magnetic flux, giving a modulation which is periodic in one flux quantum ($\Phi_0$) (Fig. 3).  We theoretically model this modulation using our numerically-computed CPR to calculate the nanoSQUID inductance ($L_S$) as a function of flux. The magnitude of $L_S$ depends on the critical current and cannot be directly measured in our capacitively-coupled resonator geometry. We thus fit the observed $f_{\mathrm{res}}$ using the geometric loop inductance and capacitance obtained from other measurements to scale the calculated CPR. The resulting fit lines are shown in Fig. 3 and describe the entire flux response very well. The values of the critical current used to generate these curves are consistent with our device geometry and our previous dc measurements of similar nanoSQUIDs.  This $L_S$ modulation measurement complements previous dc critical current ($I_C$) modulation measurements\cite{vijay_apl} of similar nanoSQUIDs.  While $I_C$ measures the maximum supercurrent that can be passed through the nanoSQUID, i.e. the peak of the SQUID's effective CPR, $L_S$ measures the inverse slope of the SQUID's CPR near zero phase.  These two quantities are not simply related in weak link junctions. In particular, since the CPR is nonsinusoidal, $I_C$ does not modulate to zero, even for an ideal SQUID with zero loop inductance, and has its minimum value at $\Phi_0/2$. In contrast, $L_S$ continues to increase ($f_{\mathrm{res}}$ decreases) past $\Phi_0/2$ provided the system remains in one of the two stable states around $\Phi_0/2$ (Fig. 3). This is technologically significant, as it implies that ac devices can be tuned over a wide range even if the dc transport properties exhibit shallow modulation. 

\begin{figure}
\includegraphics{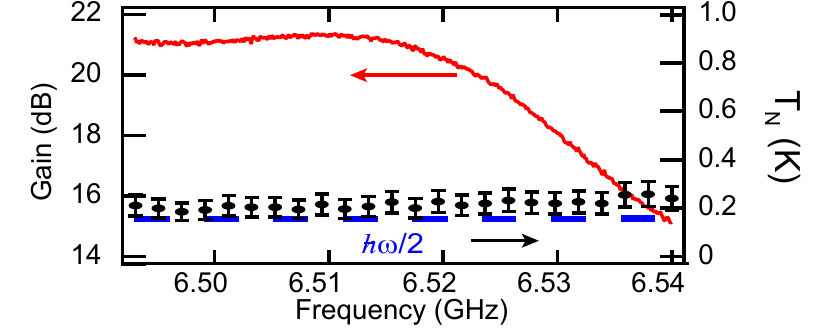}
\caption{\footnotesize\label{fig4} (color online) Amplifier performance data for the low-$Q$ resonator.  Gain (red line) of over 20dB is shown with $\approx$ 40MHz of bandwidth.  Amplifier $T_N$ (black dots)  is at or near the standard quantum limit (blue dashed line).  No spurious output tones are observed.  All data are consistent with canonical semi-classical paramp theory.}
\end{figure}

When biased into the nonlinear regime with a pump tone, our oscillators function as parametric amplifiers (paramps).  This requires drive powers just below $P_c$, the critical power at which bifurcation occurs. All our samples can be biased stably in this regime. Paramp characterization data for the low-$Q$ device is shown in Fig. 4. We measured a small signal reflected off the resonator with and without a strong pump tone; the increase in reflected signal gives the device gain.  The amplifier performance is excellent, with $>$ 20 dB of gain over $\approx$ 40MHz of bandwidth and a 1 dB compression point of -115 dBm.  Given the system noise temperature of the microwave amplification chain ($T_{sys}$), we extract the added noise of the paramp by measuring the improvement in signal-to-noise ratio (SNR) when the pump power is applied. This amplifier noise temperature ($T_N$) is plotted in Fig. 4 and corresponds to nearly one half-photon, the minimum allowed by quantum mechanics for phase-insensitive amplification\cite{caves}. The uncertainty in $T_N$, indicated by the error bars, is mainly due to uncertainty in our measurement of $T_{sys}$ using the standard hot/cold load technique.  We verified that there are no apparent spurious frequency components in the amplified output signal. A similar tunnel-junction paramp has been used in qubit readout, enabling an SNR high enough to perform continuous single-shot readout and observe quantum jumps in a macroscopic system\cite{dan_jumps}.  Our high-$Q$ resonator shows similar paramp behavior with a bandwidth roughly 100 times smaller due to its higher $Q$.  The low-$Q$ device can also be used as a high-speed, low noise magnetometer, as was done with a tunnel-junction prototype\cite{Hatridge}.  We are presently evaluating the performance of the device in this application.

These measurements directly verify the operation of 3D nanobridge junction amplifiers. The strong nonlinearity achieved suggests the potential use of nanobridges in superconducting qubits \cite{vijay_wire}. Since the critical currents of nanobridge junctions tend to be greater than a few microamps, they are more suitable for constructing phase qubits \cite{phasequbitrev,Hoskinson_camel} and we are presently exploring this application. Qubit measurements will shed light on the intrinsic losses (both low and high frequency) in nanobridge junctions and complement ongoing resonator-based measurements. We are also investigating $1/f$ flux noise in nanoSQUIDs using dispersive magnetometry techniques \cite{Hatridge}.

In conclusion, we have measured the RF response of nanobridge-based anharmonic oscillators at $T \ll T_C$.  We observe strong nonlinearity in the nanobridge CPR, enabling stable bifurcation even in low-$Q$ resonators.  These results suggest that nanobridge junctions have sufficient nonlinearity to realize superconducting qubits.  Our low-$Q$ resonator can be operated as a parametric amplifier with excellent gain and bandwidth with nearly quantum-limited noise. These results are quantitatively consistent with numerical simulations and previous dc transport measurements. We do not observe any additional loss due to the presence of a nanobridge in a resonator.  Thus, 3D aluminum nanobridge junctions are promising building blocks of superconducting amplifiers and qubits.

The authors thank Gary Williams for assistance with device imaging and J. E. Johnson for useful discussions.  Financial support was provided by AFOSR under Grant No. FA9550-08-1-0104 (R.V.) and DARPA under
Grant No. N66001-09-1-2112 (I.S.). E.M.L-F. gratefully acknowledges an NDSEG fellowship awarded by DoD, AFOSR, under Grant No. 32 CFR 168a.

\end{document}